%% file: main.tex
\documentclass{article}

\usepackage{arxiv}
\usepackage{authblk}
\usepackage{amsmath,amssymb,amsfonts}
\usepackage[utf8]{inputenc} 
\usepackage[T1]{fontenc}    
\usepackage{hyperref}       
\usepackage{url}            
\usepackage{booktabs}       
\usepackage{amsfonts}       
\usepackage{nicefrac}       
\usepackage{microtype}      
\usepackage{lipsum}
\usepackage{graphicx}
\usepackage{verbatim}

\newcommand\calW{\mathcal{W}}

\newcommand\esp[1]{{\mathchoice{\besp{#1}}{\sesp{#1}}{\sesp{#1}}{\sesp{#1}}}}
\newcommand\besp[1]{\mathbb{E}\left[#1\right]}
\newcommand\sesp[1]{\mathbb{E}[#1]}

\usepackage{enumitem}
\renewlist{cases}{enumerate}{10}
\setlist[cases]{label=\textbf{Case \arabic*}, itemindent=0pt,leftmargin=35pt,  listparindent=1.5 em}

\title{Work Stealing Simulator}

\author[1,2]{Mohammed Khatiri}
\author[1]{Denis Trystram}
\author[1]{Fr\'ed\'eric Wagner}
\affil[1]{Univ. Grenoble Alpes, CNRS, Inria, Grenoble INP, LIG, 38000 Grenoble, France}
\affil[2]{University Mohammed First, Faculty of Sciences, LaRI, 60000, Oujda, Morocco}
\affil[ ]{Email: firstname.lastname@inria.fr}  

\date{}

\begin{document}
\maketitle

\begin{abstract}
    We present in this paper a \textit{Work Stealing} lightweight PYTHON simulator.
    Our simulator is used to execute an application (list of tasks with or without
    dependencies), on a multiple processors platform linked by specific topology.
    We first give an overview of the different variants of the work stealing algorithm,
    then we present the architecture of our light Work Stealing simulator.
    Its architecture facilitates the development of other types of applications and
    other topologies for interconnecting the processors.
    We present the use cases of the simulator and the different types of results.

\end{abstract}

\input{simulator}

\bibliographystyle{unsrt}  
\bibliography{biblio}

\end{document}

%% file: simulator.tex
%

\section{Introduction}
\label{sec:system:WSS_introduction}
\subsection{Context}
The analysis of the classical \textit{Work Stealing} algorithm is a difficult combinatorial problem \cite{Blumofe1999}. 
It becomes even more difficult on more complex environments. 
For example, the analysis is much more difficult in the case on distributed memory than on shared memory since communication matter \cite{Gast2018ANA}\cite{Denis2013}.


We are interested in the analysis of \textit{Work Stealing}
algorithm on more complex environments including non homogeneous ones.
In particular, we are interested in platforms with multiple clusters where each
cluster contains a set of shared memory processors.
The clusters are linked via a not uniform interconnection network.
As processors in the same cluster communicate through a shared memory,
communications cost are almost negligible.
The processors in different clusters communicate through
the interconnection network
and thus, communications are explicit (latency or bandwidth) and costly.
\subsection{Why using simulator?}
The heterogeneity of communications and the mechanism of \textit{Work Stealing}
generate an interesting combinatorial problem,
which is more difficult than the initial case in which we use on one cluster with homogeneous communications. 
Moreover, a mathematical analysis using the potential functions is not effective as in \cite{Gast2018ANA},
because it is very difficult to find an adequate potential function.
Moreover, the worst case scenario is too far from the reality
compared to the model of \textit{Work Stealing} on one cluster.
The worst case scenario in multiple clusters is not just
when all the processors act as thieves except one,
but also when all the processors steal outside their own clusters.
This worst case scenario is one of the most difficult barrier 
to analyze the model of the \textit{Work Stealing} algorithm on multiple clusters.

Therefore, we have to rely on simulations to observe what 
happens when we use the \textit{Work Stealing} algorithm on multiple clusters platforms.
We performed simulation to understand how the algorithm behaves when the communication time increases,
and to get an idea about the average completion time
according to different parameters (communication time, number of processors, etc...).
And to compare different strategies that take 
communication time and cluster's topology into account.
At the same time, we used the simulator to validate the theoretical analysis
in the basic case of one cluster,
and show how much the Makespan is far from the experienced Makespan. 

There exist several simulators on parallel and distributed computing.
Many of them are developed for a specific research projects by researchers
and are undocumented, and/or no longer maintained.
However, there exist several High quality simulators like SimGrid~\cite{casanova:hal-01017319} that include many features and allow to consider complex situations like congestion,
cache effects for particular architectures.
However, such simulators are usually very computationally expensive, and they require a long execution time.
Our purpose is less ambitious since we target simple processing units to observe a single aspect of execution process, which is the work steeling algorithm on platforms with different topologies.
For this work, we developed a specific lightweight \textbf{PYTHON} simulator.
Our simulator is quite flexible and easy to use and update.
Moreover, it allows getting more insight on the result.
Thus, we are interested in using our own representations for interpreting the simulation results. 



\subsection{Objective}
The objective of our simulator consists in running different models of the \textit{Work Stealing} algorithm.
It executes an application on a platform,
an application consists of a list of tasks with or without dependencies,
and the platform consists of multiple processors linked by a specific topology.
The simulator allows to execute a scenario with a specific task on a specific platform.
It is designed to be sufficiently flexible to meet the different needs to analyze the
\textit{Work Stealing} algorithm and to compare different victim selection strategies.
It offers various types of applications and various topologies.
Moreover, its architecture facilitates the development of other types of applications
and other topologies for interconnecting the processors.
Even more than that, the simulator is fast. 
It also shows in details the results of each simulation.
These results could be numerical (execution time, number of steal requests, etc...) or graphical (Gantt chart, real time execution etc...).

In this paper, we give in Section~\ref{sec:system:variantsWS} an overview of the different variants of the \textit{Work Stealing} algorithm.
Then, we present in Section~\ref{sec:system:architecture} the architecture of our light \textit{Work Stealing} simulator.
In Section~\ref{sec:useofthesimulator} we use our simulator to assess the validity
of our analysis presented in \cite{Gast2018}. 
Then, we show the latency intervals exhibiting an acceptable Makespan on a single cluster,
and we conclude the section by studying the impact of simultaneous responses.



 
\section{Variants of the Work Stealing algorithm}
\label{sec:system:variantsWS}
The \textit{Work Stealing} algorithm schedules an application (set of tasks) 
in a distributed platform composed of $p$ processors linked by a specific topology.
Many algorithms and implementation variants of the Work
Stealing algorithm exist in the literature. 
In particular, we present the different task models of the scheduled application.
Then, we describe different types of platform topologies possible and how they
impact the victim selection. We conclude by describing different policies for steal answers.

\subsection{Application Task model} 
\label{sec:system:taskmodel}

The type of scheduled application is an important issue.
As see in~\cite{Gast2018ANA},
the bound of Makespan depends on the type of the scheduled application.
The application defines the characteristics of the tasks, the dependencies
between them and how the work could be divided during a steal operation.

In the literature, many researchers are interested in analyzing
\textit{Work Stealing} algorithms using different task models.
The most used task models can be classified as follows:

\subsubsection{Divisible load}
The divisible load represents applications with independent unit tasks.
It has been introduced in~\cite{Robertazzi} and experimented by \cite{Drozdowski}.
It considers the work as a divisible load where the initial amount
is represented by a single big task. Then, during execution,
each task can be divided on request into two subtasks containing each a part of its work.
For instance when a steal request occurs in a busy processor
it sends a positive response in a form of a new task containing a part of the
local work and updates accordingly its current content.
Many theoretical studies on \textit{Work Stealing} use this divisible load
model since it simplifies the theoretical analysis \cite{Denis2013}.




\subsubsection{DAG of tasks} 
This type represents an application as a set of tasks
constrained by a directed acyclic graph (DAG) of precedence \cite{COSNARDET:1993}.
This DAG has a single source that represents the first active task.
The processing time of a task can be unitary as in \cite{Arora2001} 
or depend on the size of the task \cite{Denis2013}.
The scheduling of such type is done in \cite{Arora2001},
each processor maintains a double-ended queue (called deque) of activated tasks. 
If a processor has one or more task in its deque, it executes the tasks at the head of its deque.
After completion, a task might activate other
tasks that are pushed to the end of the deque.
A task is active in the DAG only when all its precedents have been executed.
The activation tree could be a binary tree or a
fork-join whose shape depends on the execution of the algorithm.
It is a subset of the original DAG and has the same critical path.
We define the height of nodes of this tree as follows.
The height of the source as D (i.e., the length of the critical path).
The height of another task is equal to the length of its father minus one. We assume
when a processor steals work from another processor, it steals the activated tasks
with the largest height.

\subsubsection{Adaptive tasks}
The adaptive tasks represent a dynamic application that reacts specifically to the steal requests.
At the beginning, all the workload is stored as one big task which is located on a given processor.
Then during a steal operation, the processor shares a part of its task and
creates the merge task that brings together the result of the two parts at the end.
In general, the processing time of an adaptive task depends on its size and the algorithm used.
The processing time of the merge task depends on the size of the tasks that proceeded it 
and the algorithm used to merge the results.
The adaptive tasks have been studied in \cite{Daoudi2005,cung} and
introduced in \cite{Jean-Louis2006} to solve the prefix problem.

\begin{figure}[t]
	\centering
	\includegraphics[width=0.8\linewidth]{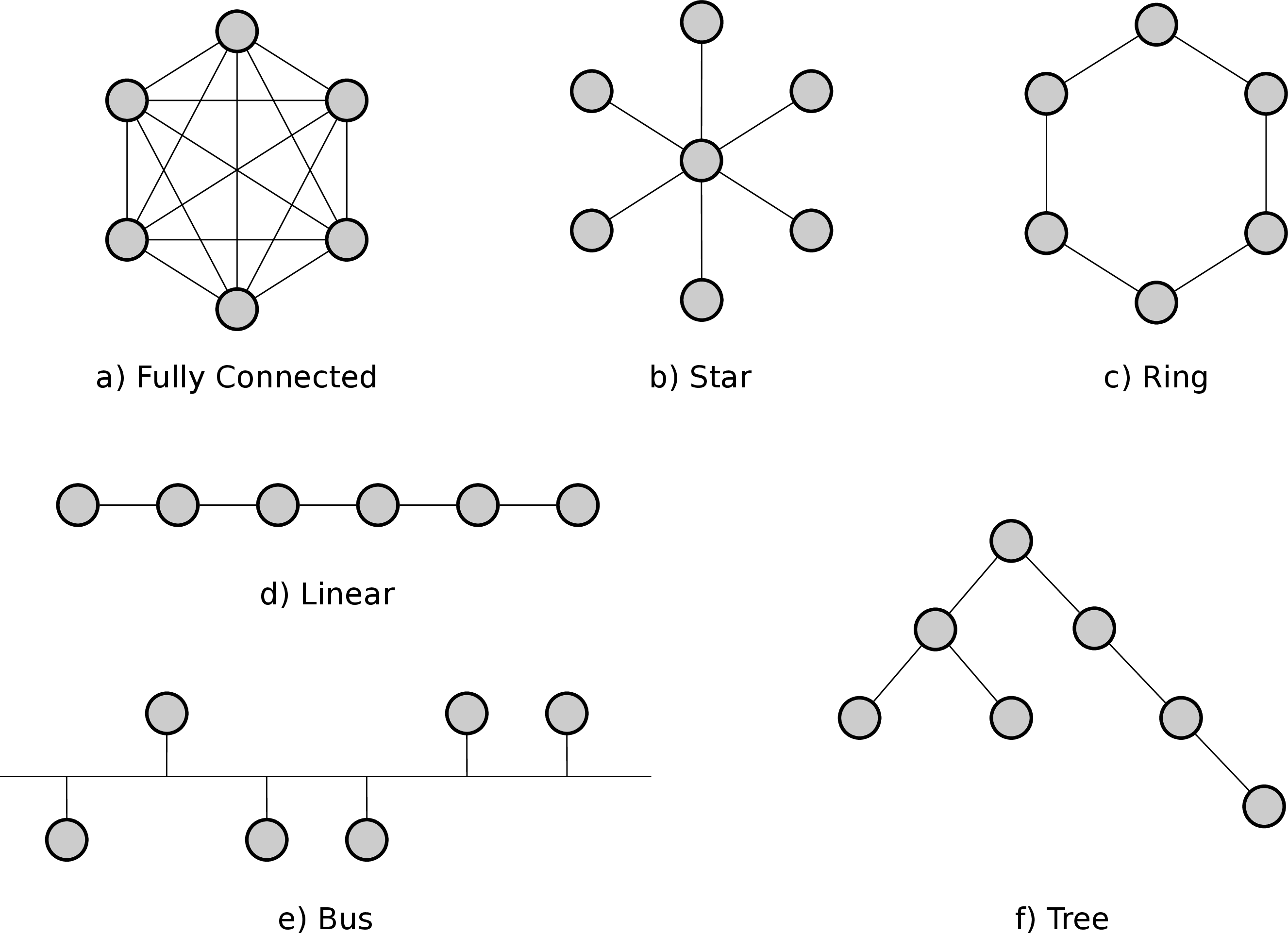}
	\caption{Multi-clusters topologies} 
\label{fig:simulator:topologies}
\end{figure}

\subsection{Platform topologies} 
\label{sec:system:platformtopology}

The platform topology defines the location of the
processors in the platform and characterizes
the communication times between them (latency or bandwidth).
There exist many topologies in the literature that can be classified as follows:

\begin{itemize}
\item \textbf{One cluster :} The same topology used in~\cite{Gast2018ANA}. 
The processors are fully connected in a cluster.
The communications between them are homogeneous and can take place simultaneously with no extra overhead, and the communication costs are dominated by the latency.
        Thus, this communication can be modeled by a constant delay (denoted by $\lambda$)
We can model the shared memory processors by a single cluster topology if we consider that the communication time takes $1$ time step.
\item \textbf{Two clusters :} the processors are divided into two clusters.
The processors in the same cluster communicate via shared memory. 
We consider that this communication takes $1$ time step.
The clusters are connected via an interconnect network that performs the communications between processors in different clusters.
Since the communication cost between cluster is much larger than the communication inside the clusters, the communication between the processors is heterogeneous and creates victim selection issue (explained in Section~\ref{sec:system:victimselection}).
\item \textbf{Multiclusters :} the processors are divided into several clusters that are linked via a network in different topologies as shown in Fig~\ref{fig:simulator:topologies}.
In these topologies, the communication between processors depends on their location and also on the location of their clusters on the topology.
\end{itemize}
\begin{figure}[t]
	\centering
	\includegraphics[width=0.8\linewidth]{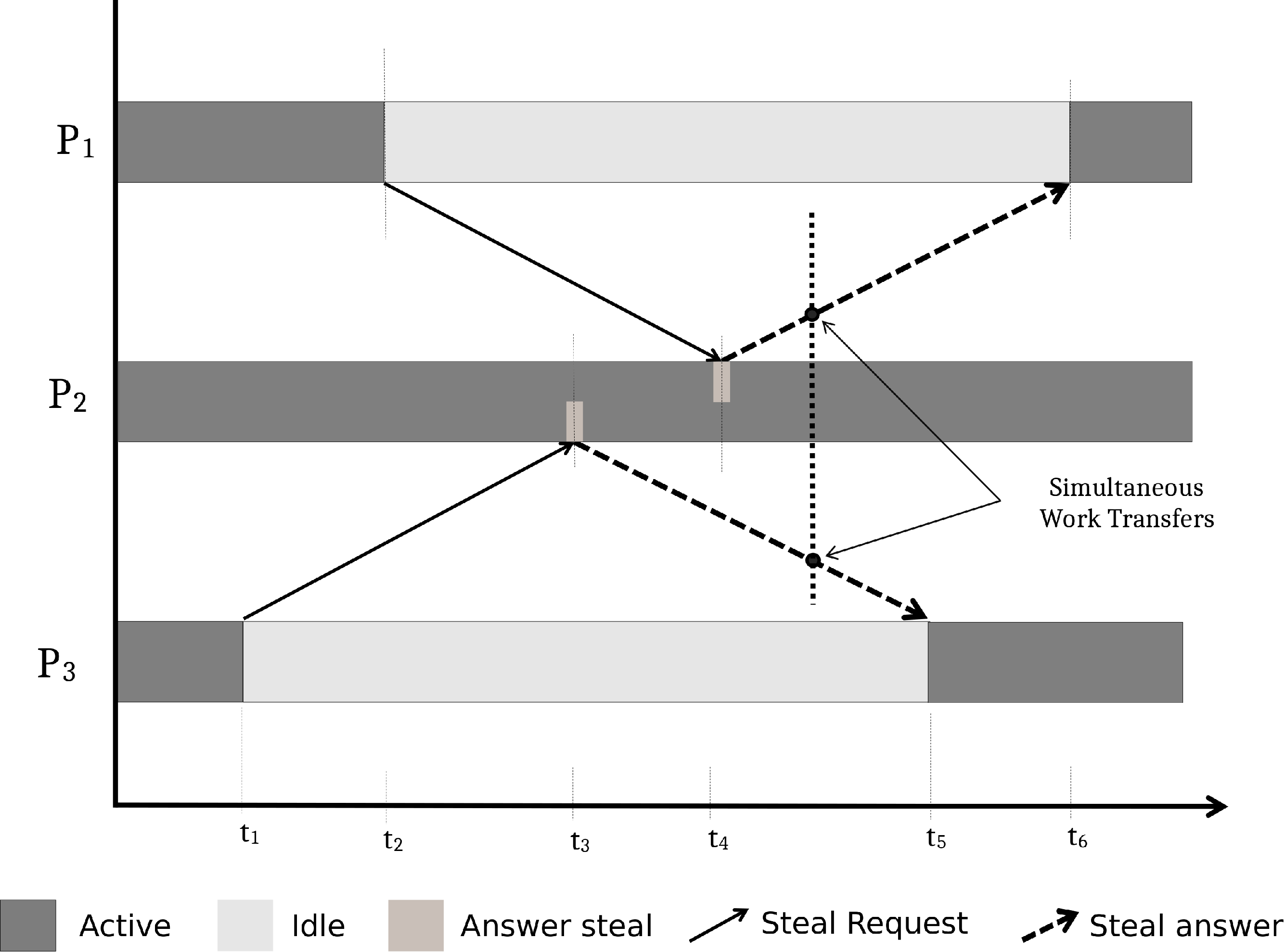}
	\caption{Distribution of work in the case of simultaneous responses} 
\label{fig:simultaneousdistribution}
\end{figure}
\subsection{Victim selection}
\label{sec:system:victimselection}
The \textit{Work Stealing} algorithm on complex topology with the heterogeneity of communication creates new questions about the victim selection strategy.
Sometimes, the victim selection should take into account the characteristic of the topology (distance between processors, the communication time, etc...).
Thus, the victim selection strategies is an important question especially on structured topology.

\subsection{Steal answer policies} 
\subsubsection{Simultaneous responses}
There exist in the literature two main variants for handling steal responses, namely, the single and simultaneous responses. 
We consider here both techniques as follows:
\begin{itemize}
	\item
	\textbf{Single work transfer (SWT)}
	is a variant where the processor can send some work to at most one processor at a time. 
	The processor sends work to a thief and it replies by a fail response to any other steal requests. 
        Using this variant the \textit{steal request} may fail in the two following cases: 
	when the victim does not have enough work or when it is already sending some work to another thief.
	\item
	\textbf{Multiple work transfers (MWT)}
	Each processor can respond and send work to several processors simultaneously. 
	The received requests are handled sequentially.
	In the classical model, the processor always answers by sending half of its work. 
	In case of simultaneous requests it arranges them in a series and answers in the same way.
    Fig~\ref{fig:simultaneousdistribution} gives an example of such simultaneous work transfers. In this figure $W_i(t)$ denotes the work on $P_i$ at time $t$.
\end{itemize}
\subsubsection{Steal Threshold}
\begin{figure}[t]
	\centering
	\includegraphics[width=0.8\textwidth]{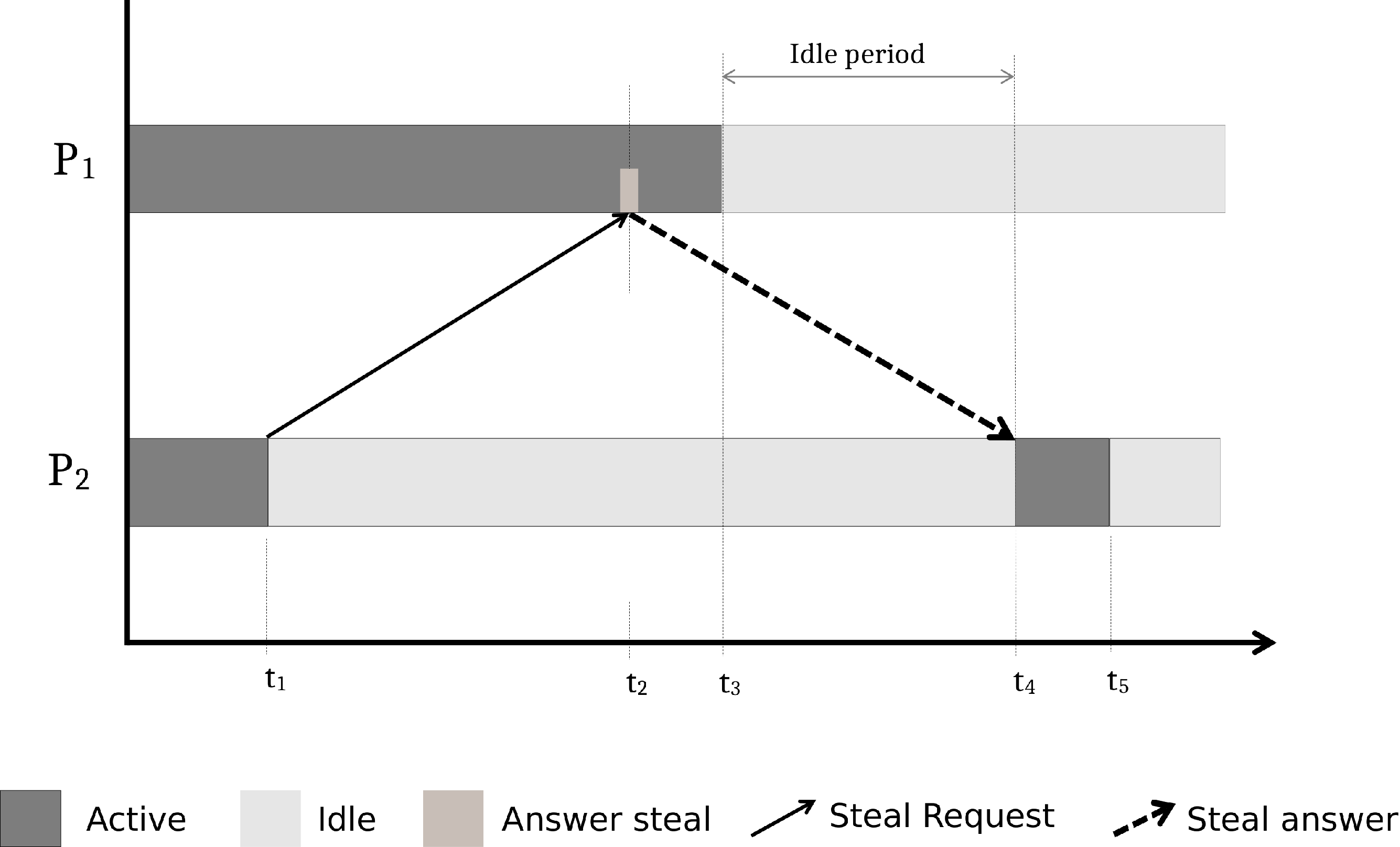}
	\caption{Example of creating artificial idle times} 
	\label{fig:stealthreshold}
\end{figure}
The main goal of \textit{Work Stealing} is to share work between processors to balance the load and speed-up the execution. 
In some cases however it might be beneficial to keep work local and answer negatively to some steal requests.

Fig~\ref{fig:stealthreshold} shows an example of this case on two processors. 
At time $t_1$ processor $P_2$ sends a steal request to $P_1$. 
At $t_2$ $P_1$ receives this request and answers by sending half of its local work, which is less than the communication duration. 
At $t_3$ $P_1$ finishes its remaining work and becomes idle. 
Then, both processors are idle in the time period between $t_3$ and $t_4$.
This clearly is a waste of resources since the whole platform is idle while there is yet some work to execute. 
Moreover, such a behavior can be chained several times. 
This effect is not purely theoretical as it has been observed during our initial experiments.

It is possible to prevent this from happening by adding a threshold on steal operations. 
We introduce a \textit{steal threshold} which prohibits steals if the remaining local work becomes too small.
 
\section{Simulator Architecture}
\label{sec:system:architecture}

We present in this section the global architecture of our simulator.
First, we describe the basic mechanism of our simulator.
Then, we explain how the simulator manages different variants of \textit{Work Stealing} (described in Section~\ref{sec:system:variantsWS}) using different independent engines. 

\textbf{Basic Mechanism.} During an execution of \textit{Work Stealing}, 
the processors switch between different states over time.
For example, a processor is active when it executes work.
Once it finishes its work, it becomes idle.
Then, if its tasks queue is not empty,
it pops a task and it becomes active again,
otherwise, it becomes a thief by sending a steal request to the other processors. 
We define an \textbf{event} as the time when a processor changes its state.
This implies that the simulator has to simulate the events time instead of all the running times continuously.
When an event occurs, the simulator uses the model instructions to execute it.
For example, when a processor becomes a thief,
the simulator chooses the victim using the strategy defined by the considered model and sends a steal request to the selected victim.

The execution of a simulation returns different statistical results
(simulation time, number of steal requests, etc...).
Other type of results are possible, for example,
we can generate some logs to show the Gantt execution chart.
We can also display the DAG execution which delivers the execution in real time.

In our work, the simulator is used to experimentally analyze different variants of \textit{Work Stealing}.
Thus, it should handle the different variants described in Section-\ref{sec:system:architecture}.
Moreover, the simulator needs to manage the different types of application, the different topologies and all other variants.

For all these reasons, 
our simulator\footnote{git@github.com:mkhatiri/ws-simulator.git} is designed to be sufficiently flexible in order 
to simulate different \textit{Work Stealing} models.
Its flexibility aims to allow us to experiment with different \textit{Work Stealing} algorithms,
different topologies, different steal strategies and different types of application.
The simulator should also generate a sufficient amount of logs 
for a detailed analysis each tested scenario.

We decompose the simulator into several independent engines.
Each engine develops a part of the simulator 
and offers an operating interface which presents the main provided functionalities. 
The engines interact between them through these operating interfaces.

\begin{figure}[htb]
	\centering
	\includegraphics[width=0.8\textwidth]{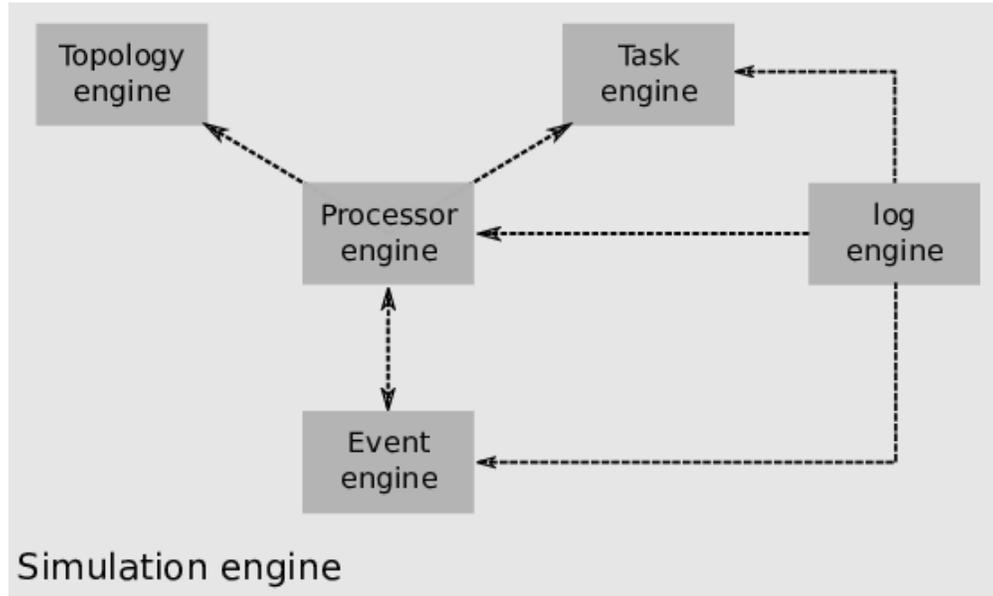}
	\caption{The different engines of our simulator}
	\label{fig:simulator:simulatorArchitecture}
\end{figure}

The overall architecture of our simulator is composed of six main engines,
as seen on Fig~\ref{fig:simulator:simulatorArchitecture}.
The event engine is the core of our simulator,
it manages the processors events during the time
to run the simulation of a scenario.
The events are executed through the processor
engine which provides different functionalities to perform the \textit{Work Stealing} algorithm. 
The processor engine uses the task engine to manage the execution of tasks and uses the topology engine to manage the interactions
between the processors.
During the execution of a simulation, the log engine keeps track of different information and generates different logs.
The rest of this section details the role of each engine
and explains the interactions between them.

\subsection{Event engine}

The event engine represents the kernel of the simulator.
In this section, we first explain the global idea to run the simulation of a scenario.
Then we define the components used by the event engine
to simulate an execution of an application defined by task engine
on a platform defined by topology engine.

\begin{figure}
	\centering
	\includegraphics[width=0.8\textwidth]{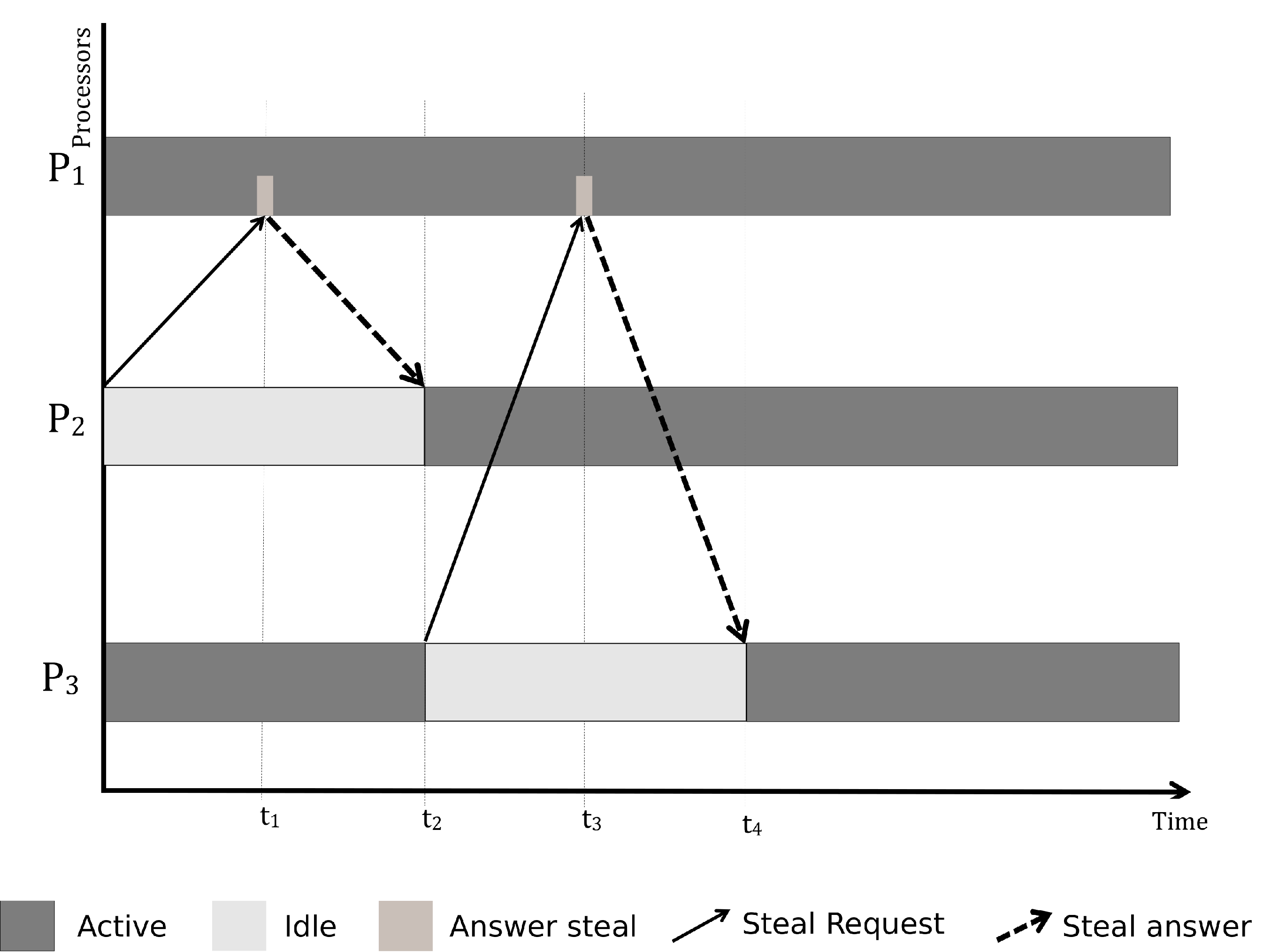}
	\caption{Example of a \textit{Work Stealing} execution}
	\label{fig:simulator:ws}
\end{figure}

In the \textit{Work Stealing} algorithm, a processor switches between different possible states.
Fig~\ref{fig:simulator:ws} presents an example of \textit{Work Stealing} execution,
each processor interacts when it becomes idle ($P_3$ at $t2$),
when it receives a steal request ($P_1$ at $t_3$)
or when it receives a steal answer ($P_3$ at $t_4$).


The global idea of our simulator consists in simulating
a set of discrete \textbf{events} through time instead of simulating the whole execution time,
where an \textit{event} stands for changing the state of a processor at a specific time.
For that, the event engine lists the available events on a heap
and executes them sequentially according to their time.
The execution of events follows different steps to update the system and creates new events in the global heap, these events will be executed following the same mechanism.

We define an event by its time, its related processors and its type.
Based on the different states of a processor.
We consider the three types of possible events : 
\begin{itemize}
	\item \textbf{Idle event}: a processor finishes its running task. 
	 When a processor has an idle event, it means that it is executing a task.
	 Thus, the time of this event is defined by the execution time of the related task.
	 
	\item \textbf{Steal request event}: a processor receives a steal request.
	\item \textbf{Steal answer event}: a processor receives the answer after a steal request.
\end{itemize} 

The event engine offers two functions to manage the event heap,
\textit{next\_event()} which pops the nearest event from the global events heap 
and \textit{add\_event()} which adds an event to the global event heap.
The event engine controls also the global simulation time which starts at $0$.
All tasks type described below start with one big task.
Thus, at the beginning of the simulation, 
the first processor executes the first task of the application, then it starts the simulation with the related \textit{idle event}.
All other processors start with an \textit{idle event} that occurs at the beginning of the simulation (time $0$).
The event engine starts simulation with a global event heap that contains all the first events.

To run a simulation, the event engine call \textit{next\_event} to get the nearest event,
then it updates the global simulation time according to this event time,
and then it executes this event.
The same processes will be used for other events.
The event engine uses the task engine to detect the end of the simulation.
(We will detail that in Section-\ref{sec:system:tasksEN} ).
The execution of an event interacts on the related processors and orders it to update its state and creates other events.
The execution time is defined by the last executed event. 

Before explaining the processor engine which performs the execution of the events.
We present the task engine and topology engine that will be used extensively by the processor engine.

\subsection{Task engine}
\label{sec:system:tasksEN}

The main objective of our simulator is to simulate the performance of different variants of the \textit{Work Stealing} algorithm.
The first variant consists in managing different types of applications.
Where an application is defined by a set of tasks with or without precedence constraints.
The task engine is used to handle everything related to the application during a simulation.

As stated, an application could be modeled as a divisible load or application with adaptive tasks.
In these two types, the task could be divided during a steal request, 
Moreover, the adaptive task split the task into two subtasks and generates the merge task which depends on these two subtasks.
Therefore, our idea is to define a method to \textbf{split} work during a steal request.
Then, each application type defines this function according to its characteristics.
For instance, the split function in application with divisible load divides a task into two subtasks.
In case of application with DAG of task where the steal is handled from the processor queue. Then, the split function return \textbf{None} since the tasks can not be splitted.  

The execution of a task may activate one or more tasks as in case of DAG task or the application with adaptive tasks, where the execution of task may active the merge task if it exists.
To manage this, the task engine defines a method to 
update the task dependencies when a task is completed.

For all these reasons, the task engine provides an operating interface which offers all the needed functionalities to manage tasks. It also controls the global application.
Then, the implementation of a new type of application simply requires the redefinition of the operating interface functions.

We first describe what is needed to manage a task during an execution.
Task management consists of controlling the execution time of each task, updating the dependencies when finishing the execution of a task, and splitting tasks between two processors during a steal request.
Thus, the operating interface of task engine is based on the following functions: 

\begin{itemize}
\item  \textbf{init()} : used to create a new task during a simulation. 
\item  \textbf{split()} : used to split the task during a steal and returns \textbf{Non} if the task can not be divided..
\item  \textbf{end\_execute\_task()} : used to update dependencies when a task is completed. 
\item  \textbf{get\_work()} : used to compute the execution time of the task. 
\end{itemize}
The task engine offers the mechanism to detect the end of an execution.
It uses two global variables, one to compute the number of created tasks in the system (updated each init() call)), and to compute the number of completed tasks (updated each \textit{end\_execute\_task()} call).
The execution finishes when the created tasks are equal to the completed tasks.

To simplify the simulation, the task engine offers different functions that automatically generate different application based on DAG tasks.
It also offers a function to use a predefined application as input. 
For this, the predefined application must be described in JSON format
that defines the tasks logs (Section~\ref{sec:system:LogEN}).


%
\subsection{Topology engine}
\label{sec:system:topologyEN}

We target to simulate the \textit{Work Stealing} algorithm on platforms with different topologies.
A topology defines the distribution of the processors on the platform and the communication characteristics between them.
We explain in Section~\ref{sec:system:platformtopology} the different type of topology.
The topology engine is used to manage different platform topologies. 

To simulate \textit{Work Stealing} algorithm, the topology is used for knowing the communication time between two processors during a steal operation. Moreover, since the victim selection strategy depends on the processor topology, the topology engine is also used to manage different victim selection strategies.
Thus, the engine defines the function \textit{distance()} which returns the communication between two processors in the platform, and the \textit{select\_victim()} function which return the id of another processor based on specific strategy.

The topology engine is also used to manage different parameters
which are used by the \textit{Work Stealing} algorithm during an execution, 
for example, \textit{is\_simultaneous} is used to
determine if a processor can send work to several processors at the same time. The mechanism used to manage this option is defined by the processor engine. 
It also defines \textit{steal threshold} parameters which can be static or depend on the communication time.

\begin{figure}[htb]
	\centering
	\includegraphics[width=0.8\textwidth]{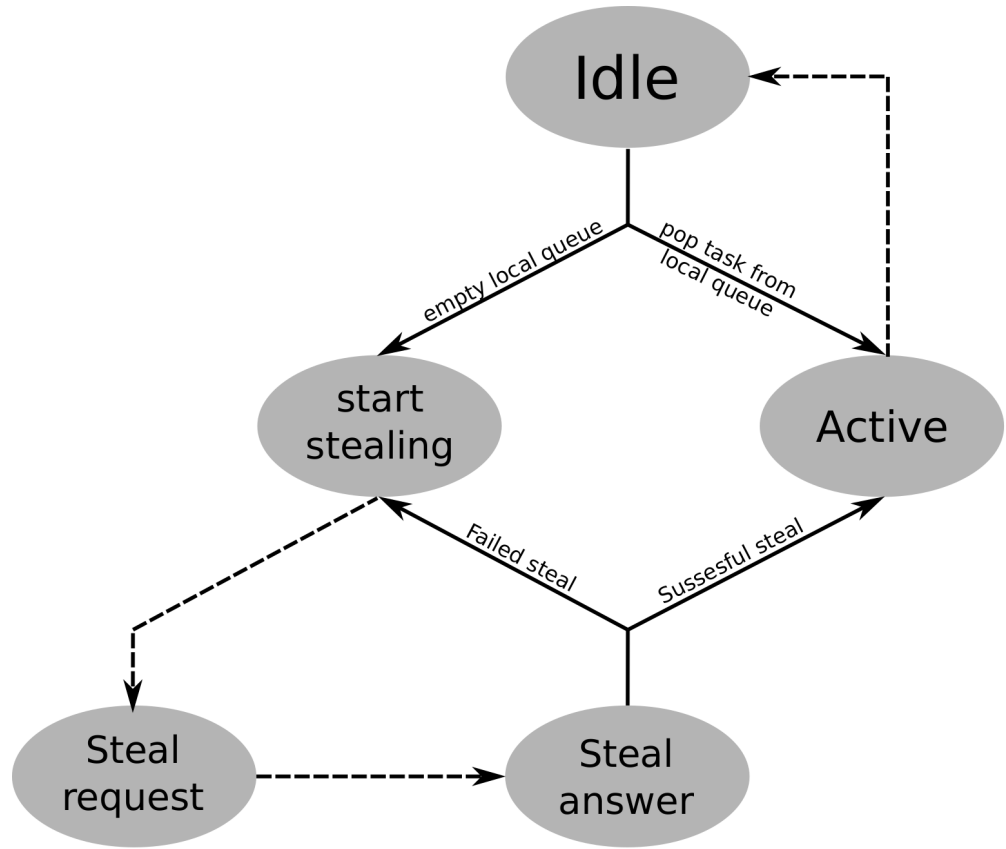}
	\caption{States cycle of a processor}
	\label{fig:simulator:ProcesorState}
\end{figure}

\subsection{Processor engine}
\label{sec:system:ProcessorEN}

The processor engine manages the processors state on the simulator,
it offers all necessary functionalities to update a processor when a related event occurs.
These functions apply the mechanism of \textit{Work Stealing} algorithm.

The process followed by a processor during the execution
of \textit{Work Stealing} algorithm is defined as shown in Fig~\ref{fig:simulator:ProcesorState}.
An active processor becomes idle when it finishes its running task.
An idle processor becomes active if it finds tasks in its local queue,
otherwise, it becomes a thief and it sends a steal request
to another processor (called victim).
Once the victim receives the request (\textit{steal request event}), it answers by some of its tasks or failed. 
Once the thief processor receives the answer (\textit{steal answer event}), it becomes active if the steal succeeds, or it becomes a thief again if the steal failures.

The processor engine provides for each processor different methods to process the \textit{Work Stealing} algorithm. These functions are organizing as follows:

\begin{itemize}
	\item \textbf{idle()} : used when a processor finishes its running task. Its main steps are as follows :
	It uses task engine to call \textit{end\_execute\_task()} for the finished task. This operation may active other tasks on the processor local queue.
	
	Then, the function checks the processor local queue, if is not empty, the processor pops a task from it,
							and creates the \textit{idle event} correspond.
							Otherwise, the processor performs a steal request (by calling the \textbf{start\_stealing()} function).
							
	\item \textbf{start\_stealing()} : used to perform the steal operation. This operation requires a victim selection,
										and produces a \textit{steal request event}.
										The victim selection is issued by topology engine (by calling \textbf{select\_victim()}).
										Once the victim is selected, 									   
									    then, it computes the communication time between
									    to send the request to the selected victim,
									    and finally it creates the corresponding \textit{steal request event}.
									   
	\item \textbf{answer\_steal\_request()} : used when a \textit{steal request event} occurs.
	It performs the answer operation.
	An answer response moves (if it is possible) the work from the victim to the thief.
	
	In this function, \textbf{get\_part\_of\_work\_if\_exist()} is used to compute the stolen tasks.
											  
	The steal failed in two cases : if there is no work to share, 
								  or if the processor is already busy with another steal answer and the topology does not allow simultaneous answer.
											  
	Once the stolen task is ready, this function uses topology engine to compute the communication time to answer this request in order to create the corresponding \textit{steal answer event}.
	
	\item \textbf{get\_part\_of\_work\_if\_exist()} : used to compute the stolen task. The processor checks its tasks queue, if it is not empty, this function returns a task from it,
	otherwise, the processor tries to split its running task using the \textbf{split()} function defined by the task engine,
	if the current task is split. It updates the \textit{idle event} correspond to the running task before splitting.
	
	\item \textbf{steal\_answer()} :  used to trait the answer request which contains the stolen task. Two cases are possible, if the stolen task contains work, it creates the \textit{idle event} corresponding the execution of the stolen task, Otherwise, the processor will try to steal work again by calling the \textbf{start\_stealing()} function.
\end{itemize}

These functions are used by the event engine to execute the three event types as follows:

\begin{itemize}
\item The execution of an \textit{idle event} call the \textbf{idle()} function.
\item The execution of a \textit{Steal Request Event} uses the victim to call the \textbf{answer\_steal\_request()} function.
\item The execution of a \textit{Steal Answer Event} use the thief to call the \textbf{steal\_answer()} function.
\end{itemize}



\subsection{Log engine}
\label{sec:system:LogEN}

The simulator is used to experimentally analyze different models of the \textit{Work Stealing} algorithm.
It should therefore generate sufficient results that simplify the analysis of the execution of a scenario.
For this reason, the log engine is used to provide different functionalities to keep trace of different information during the execution.

Several pieces of information are needed to analyze the execution of a scenario.
For instance, we need the global execution information such as execution time and the number of steal requests.
These results are presented in digital format.
Other information are useful like the processes state over the whole execution or the final shape of the application executed, thus, the engine should log the different changes on the application and on the processor states and their interaction during the simulation.

For these reasons, the log engine uses processor engine and task engine functionalities to keep track of simulation information.
For instance, to account the global number of steal request, the log engine initialize the number of steal requests to $0$ at the beginning and increments it each time the \textbf{Answer\_steal\_request()} function is called.
In another example, the log engine captures the dependencies each time the \textbf{split()} defined in the task engine is called.


After a simulation, the overall results like (execution time, number of successful and failed steal requests, total work executed, etc...) are displayed in the console in digital format.
Moreover, the simulator offers the possibility to generate other special logs that can be transformed into graphic format using standard trace analysis tools (\textit{Paje} file format \cite{Paje2000} and \ref{OutilsDeFred}).

\begin{figure}[!]
	\centering
	\includegraphics[width=0.8\textwidth]{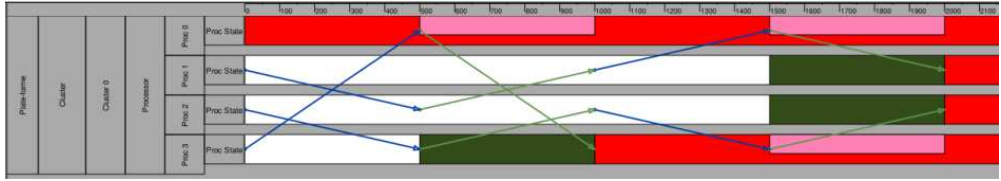}
	\caption{Gantt Chart of the first step of the execution}
	\label{fig:simulator:GantCharttbeginning}
\end{figure}

For instance, Fig~\ref{fig:simulator:GantChartt} depicts the Gantt chart of the processors during the execution simulation of a scenario generated by our simulator, and displayed using \textit{Paje}.
Through this presentation, we can analyze and understand what happened in the whole execution or a part of it.
For instance, we can focus on the first phase of the execution to understand how the work is distributed as in Fig~\ref{fig:simulator:GantCharttbeginning}.

\begin{figure}[!]
	\centering
	\includegraphics[width=0.9\textwidth]{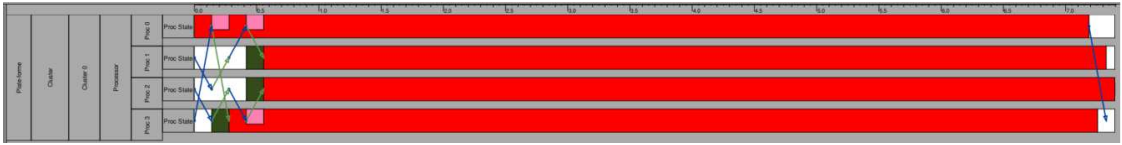}
	\caption{Gantt Chart of the whole execution}
	\label{fig:simulator:GantChartt}
\end{figure}

\begin{figure}[htp]
	\centering
	\includegraphics[angle=90,width=0.78\textwidth]{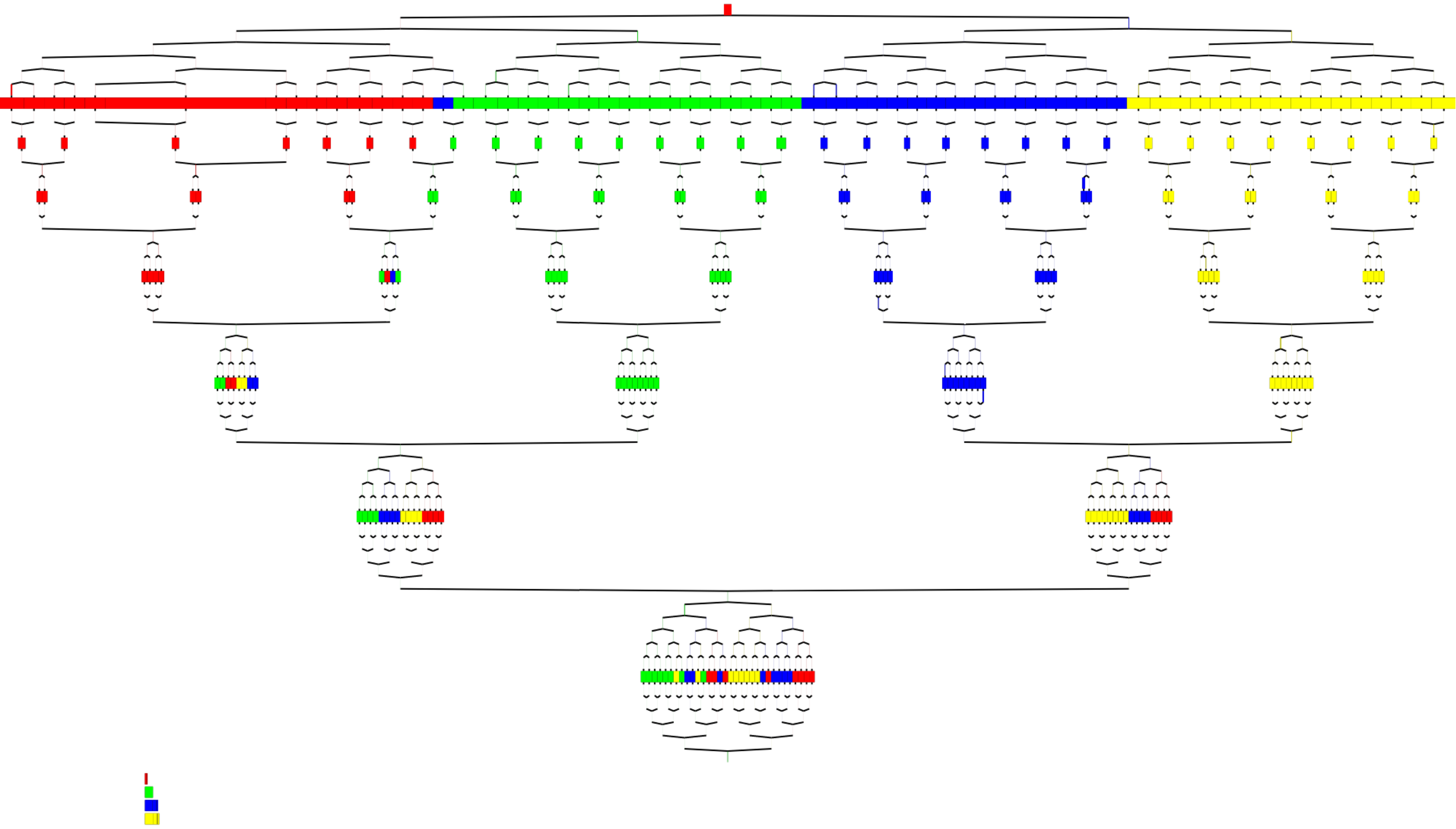}

	\caption{Execution graph of a DAG task application (Merge sort)}
	\label{fig:simulator:simulationmergesort}
\end{figure}

The simulator offers also the possibility to generates the executed application as output file with a \textit{JSON} format.
The \textit{JSON} file store for each task of the application all information as the dependencies, the work , the start and finish execution time and the processor that executes it.
The JSON file can be displayed using a \textit{JSONTOSVG} tools developed by Frederic Wagner in \cite{Outils_De_Fred}.
Fig~\ref{fig:simulator:simulationmergesort} depicts the execution graph of an application scheduled by our simulator. 
The colors present the processor, there are useful for understanding the impact of steals on the execution processes.

\subsection{Simulator engine}
\label{sec:system:CentralEN}

A simulation requires several configurations.
We need to initialize and configure the application and the platform with its topology.
Then, we need to configure different variants of \textit{Work Stealing} algorithm.
The Simulator engine is used to gather all engines to perform the different initialization before starting a simulation. 

The principle of the experimental analysis is to obtain several execution results for a scenario in order to analyze the average or the limits.
To analyze the impact of a variable, we need to simulate different scenarios for this variable to plot the results according to this variable.
For instance, to analyze the impact of the communication latency on the Makespan, we need to run several scenarios for different value of the communication latency, then we analyze the average makespan according to latencies.

For these reasons, the simulator engine proposes for the users a control panel
which allows the possibility to configure the different parameters of a scenario as the application and the platform.
It allows the user to configure the number of executions for each scenario. 
It also allows the user to set the interval of configuration values.
The simulator engine is developed to run several scenarios and simulation in the same time.
This option allows users to save the execution time by running several experiments in common.


\section{Use of the simulator}
\label{sec:useofthesimulator}

\subsection{Validation and discussion of the theoretical analysis}
\label{sec:analysisvalidation}

In our paper~\cite{Gast2018ANA}, we proved a new upper bound of the Makespan of the \textit{Work stealing} algorithm with an explicit latency on a one cluster topology.
The objective of this section is to use the simulator to experiment the \textit{Work stealing} algorithm in order to confirm and discuss the theoretical results and to refine the constant $\gamma$.


\subsubsection{Configurations}
We configure our this simulator to follows the model of independent
tasks described in \cite{Gast2018ANA} to schedule
$\mathcal{W}$ unitary independent tasks on a distributed platform
composed of $p$ identical processors in one cluster topology.
Between each two processors,
the communication cost is modeled by a constant delay represents the latency.
(denoted by $\lambda$)


Each simulation is fully described by three
parameters: $(\calW, p, \lambda)$.
For our tests, we vary the number of unit tasks $\mathcal{W}$ between $10^5$ and $10^8$, the number of processors $p$ between 32 and 256 and the latency $\lambda$ between 2 and 500.
Each experimental setting has been reproduced 1000 times in order to compute median or interquartile ranges. 

\subsubsection{Validation of the bound and definition of the ``overhead ratio''}

As seen before, the bound of the expected Makespan consists of two
terms: the first term is the ratio $\mathcal{W}/{p}$ which does not
depend of the configuration and the algorithm, and the second term
which represents the overhead related to work requests.
\begin{align*}
    \qquad &\esp{C_{max}} \leq \frac{\mathcal{W}}{p} +  4\lambda\gamma\log_2\frac{\mathcal{W}}{\lambda} 
  \end{align*}
Our analysis bounds the second term to derive our bound on the Makespan.
To analyze the validity of our bound, we define what we call the
\emph{overhead ratio} as the ratio between the second term of our
theoretical bound ($4\gamma\lambda\log_2(\mathcal{W}/\lambda)$) and
the execution time simulated minus the ratio $\mathcal{W}/p$:
for a given simulation, we define
\begin{align*}
  \text{Overhead\_ratio}=\frac{4\gamma\lambda\log_2(\mathcal{W}/\lambda)}
  {\text{Simulation\_time}  - \frac{\calW}{p}}
\end{align*}
We study
this overhead ratio under different parameters $\mathcal{W}$, $p$ and
$\lambda$. 
\begin{figure}[!]
  \centering
  \includegraphics[width=0.7\linewidth]{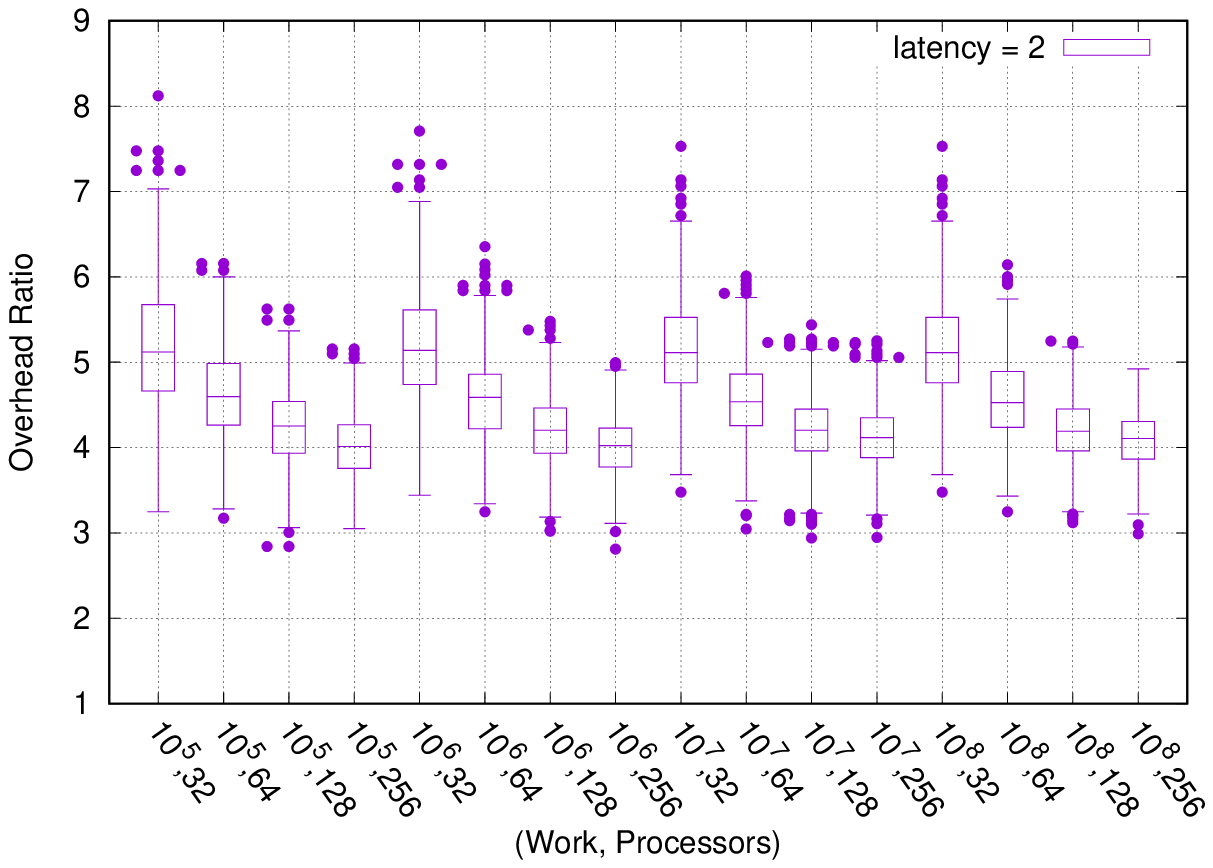}

  \includegraphics[width=0.7\linewidth]{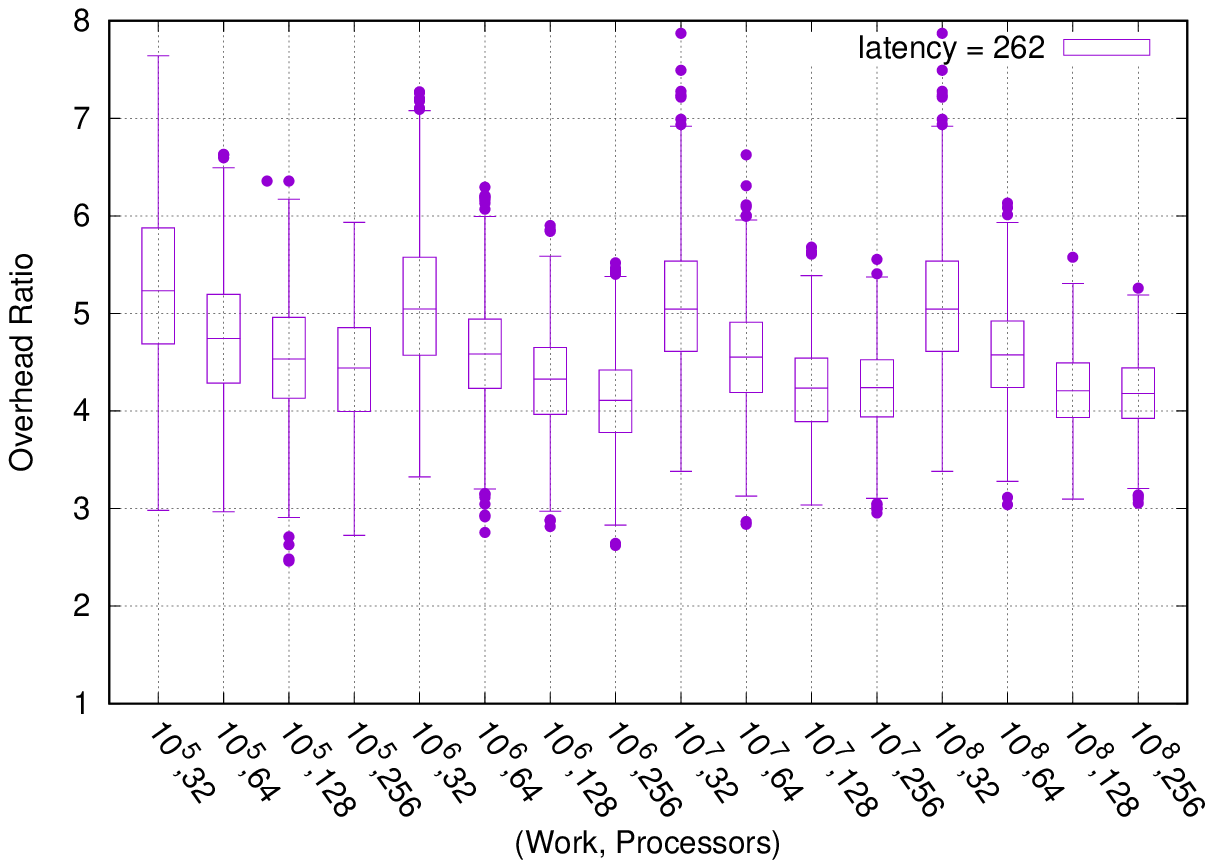}

  \includegraphics[width=0.7\linewidth]{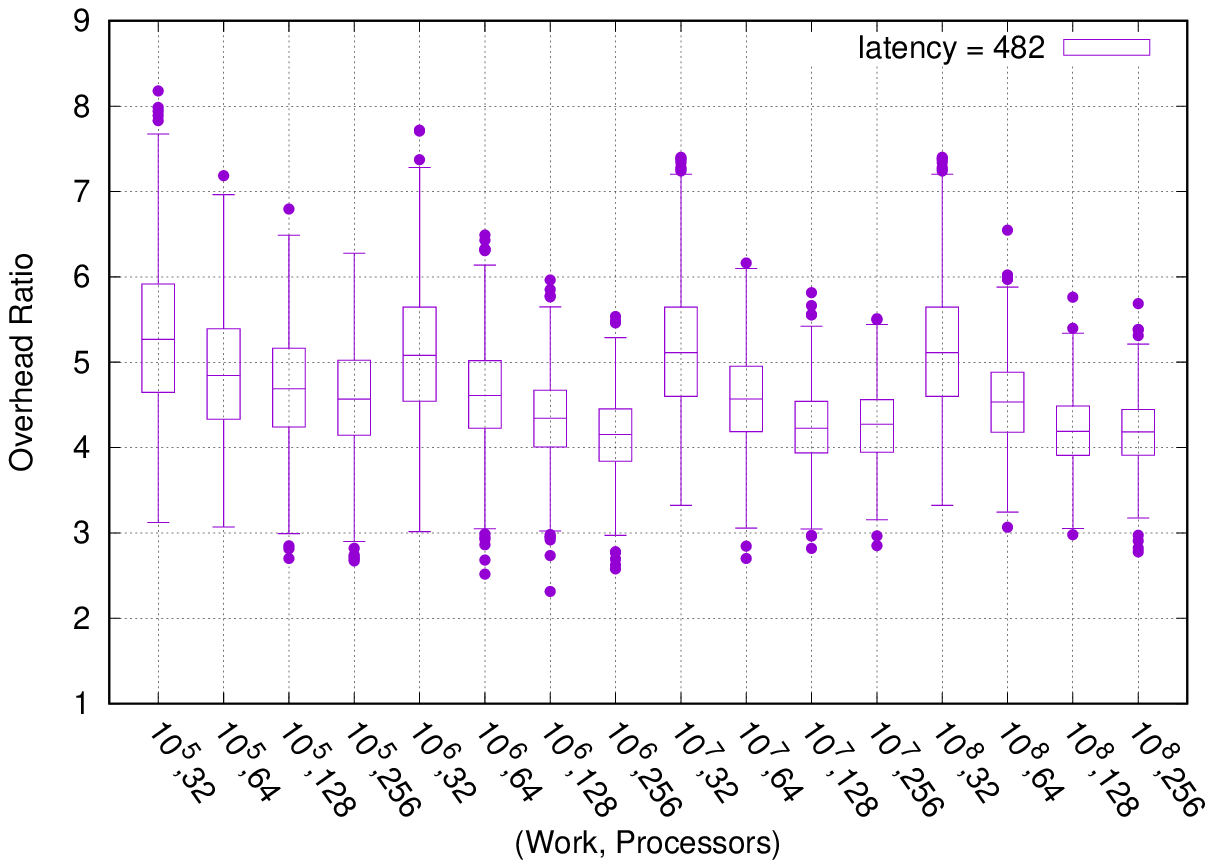}
  
  \caption{Overhead ratio as a function of $(\mathcal{W},p)$ for different values of latency $\lambda$}
  \label{fig:accuracy}
\end{figure}

Fig~\ref{fig:accuracy} plots the overhead ratio according to each
couple $(\mathcal{W}, p)$, for different latency values  $\lambda=\{2,\ 262,\ 482\}$ units of time. 
The~x-axis is $(\mathcal{W}, p)$ for all values of
$\mathcal{W}$ and $p$ intervals and the y-axis shows the overhead
ratio.  We use here a BoxPlot graphical method to present the
results. BoxPlots give a good overview and a numerical summary of a
data set.  The ''interquartile range'' in the middle part of the plot
represents the middle quartiles where 50\% of the results are
presented.
The line inside the box presents the median. The whiskers
on either side of the IQR represent the lowest and highest quartiles
of the data.  The ends of the whiskers represent the maximum and
minimum of the data, and the individual points beyond the whiskers
represent outliers.

We observe that our bound is systematically about 4 to 5.5 times greater
to the one computed by simulation (depending on the range of parameters).
The ratio between the two bounds decreases with the number of processors but seems fairly independent
to $\calW$.

\subsubsection{Discussion}

The challenge of this work is to analyze the \textit{Work Stealing} algorithm with an
explicit latency.
We presented a new analysis which derives a bound
on the expected Makespan for a given $\mathcal{W}$, $p$ and~$\lambda$.
It shows that the expected Makespan is bounded by $\calW/p$ plus an
additional term bounded by $4\gamma\approx16$ times
$\lambda\log_2(\calW/(2\lambda))$.
As observed in Fig~\ref{fig:accuracy}, the constant $4\gamma$ is about four to
five times larger than the one observed by simulation.
A more precise fitting based on simulation results leads to the expression
$\mathcal{W}/p + 3.8\lambda\log_2(\mathcal{W}/\lambda)$ (the value
$3.8$ is a fitting computed on all our experiments).
The discrepancy between the theoretical bound of $16$ and the experimental result of
$3.8$ essentially comes from the different approximations that were done in the proof.
%
$h(r)$ of Case~2.  This analysis could probably be improved by taking
%
%
%
%
%
%
\subsection{Acceptable latency}

The combination between the theoretical bound and the experiment fitting of the constant lend to the Makespan analytical expression $\mathcal{W}/p + 3.8\lambda\log_2(\mathcal{W}/\lambda)$.
One of the first uses of this expression is to predict when a given $\frac{W}{p}$ and $\lambda$ configuration will yield acceptable performances. 
Using the Makespan expression we observe that two parameters dominate: The $\frac{W}{p}$ ratio in the first term and $\lambda$ which impacts the second term of the formula representing the overhead due to communication delays. 

As stated before $\frac{W}{p}$ is a good lower bound on the best possible Makespan. A Makespan $C_{\max}$ is acceptable if the ratio $C_{\max}/C_{\max}^*$ is close to 1, where $C_{\max}^*$ is the best possible Makespan.
In our analysis, we consider a Makespan $C_{\max}$ as acceptable if $\frac{C_{\max}}{(W/p)} \leq 1.1$ (overhead less than 10\%).
We study here which configurations allow us to obtain such an acceptable Makespan.
Using the time estimation Formula we derive the equation below linking $W$, $\lambda$ and $p$
in order to get an acceptable Makespan.

\begin{center}
	$\frac{W}{p} + 3.8\log_{2}(\frac{W}{2\lambda})\lambda = 1.1\frac{W}{p} $
\end{center}
\begin{figure}[h]
	\centering
	\includegraphics[width=0.8\linewidth]{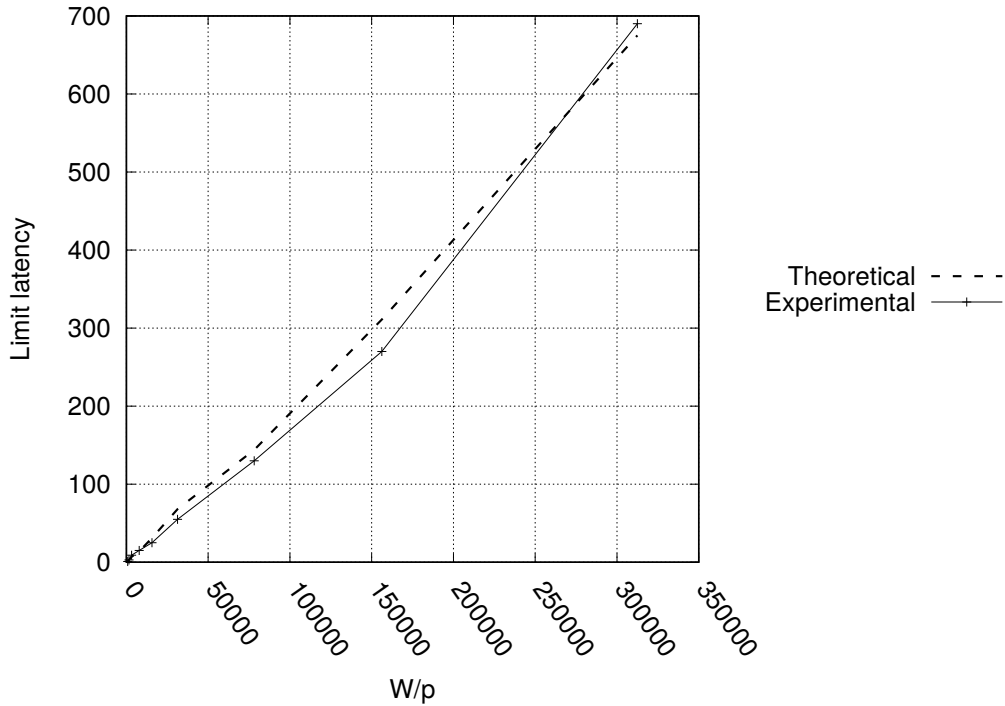}
	\caption{Limit latency exhibiting an acceptable Makespan according to $\frac{W}{p}$} 
\label{fig:latency_interval}
\end{figure}
Using this equation we can easily predict when a given $W$,  $p$ and $\lambda$ yields acceptable performance. Moreover for a specific $W$ and a fixed $\lambda$ we can easily choose the maximum number of processors applicable.

To verify the validity of this formula we solve numerically this equation for different $\frac{W}{p}$ to get the theoretical limit latency for an acceptable Makespan. 
We then verify experimentally the obtained solutions. So for a fixed  $W$ and $p$ we test different $\lambda$ and take the maximal one yielding an acceptable Makespan. We call this \textit{the experimental limit latency}. With this result we are able to compare the theoretical and the experimental limit latency.
Fig~\ref{fig:latency_interval} plots the theoretical and experimental limit latency according to $\frac{W}{p}$. The x-axis is $\frac{W}{p}$ for $W$ between $10^5$ and $10^8$ and $p$ between 32 and 256 y-axis show the limit latency.

In Fig~\ref{fig:latency_interval} we observe that the two curves overlap and conclude again on the good accuracy of our prediction. 
Moreover we can see that the relation between the latency limit and the $\frac{W}{p}$
ratio is close to linear. Using this figure we can derive the following equation:
$\frac{W}{p} = 470\lambda$. Using this equation it is easy to evaluate performances for a given $W$, $p$ and $\lambda$. In addition it allows us to compute easily for any configuration the maximal number of processors $\frac{W}{470\lambda}$ yielding an acceptable Makespan.
\begin{figure}[htp]
	\centering
	\includegraphics[width=0.8\linewidth]{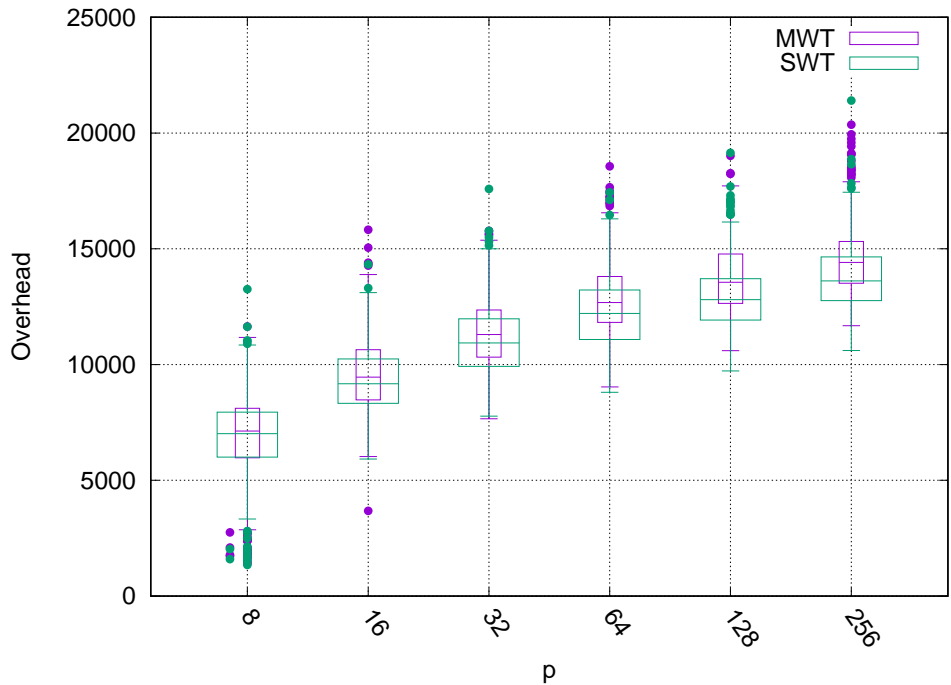}
	\caption{The Overhead of the execution using \emph{MWT} and \emph{SWT} according to the number of processors ($\lambda=262$ and $W=10^8$)} 
\label{fig:overheadsimultaneousefig}
\end{figure}
\subsection{The impact of simultaneous responses}
\label{Sec:simultaneousresponse}
We use in this section our simulator to study the influence of the multiple
work transfers mechanism (\emph{MWT}) on One cluster topology.
In our experimental runs, we compare the results obtained using both variants: With multiple work transfers and with a single work transfer (\emph{SWT}).
\begin{figure}[htp!]
	\centering
	\includegraphics[width=0.9\textwidth]{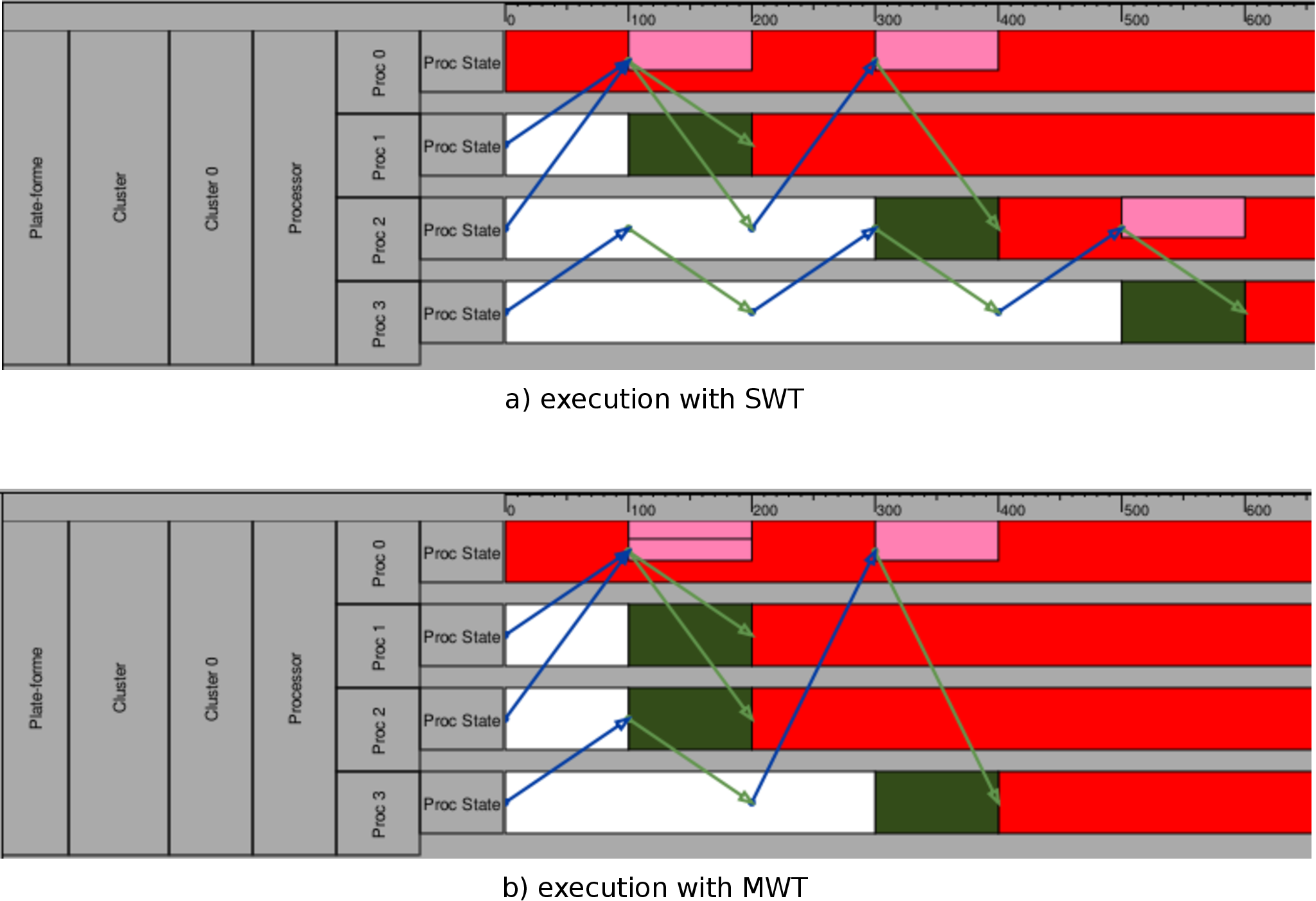}
	\caption{Gantt chart of the first phase of execution, comparison between \emph{MWT} and \emph{SWT}}
	\label{fig:simulator:GantChartt}
\end{figure}
Fig~\ref{fig:overheadsimultaneousefig} depicts a comparison between LWR and SWR showing the overload obtained using each mechanism according to the processor number.
This show that the \emph{MWT} mechanism does not bring a significant gain in the overall performances, which spurred us to analyze in detail the execution traces. In this analysis we remark that any execution using a \textit{Work Stealing} algorithm decomposes into three phases. The first phase which is denoted by the startup phase, when all the processors try to have work. This phase finishes when all processors become active. The second phase corresponds to the situation in which all processors have work and just a few steal requests between processors happen.
The last phase starts when there is little work and the majority of processors are inactive.  
\begin{figure}[ht!]
	\centering
	\includegraphics[width=0.8\linewidth]{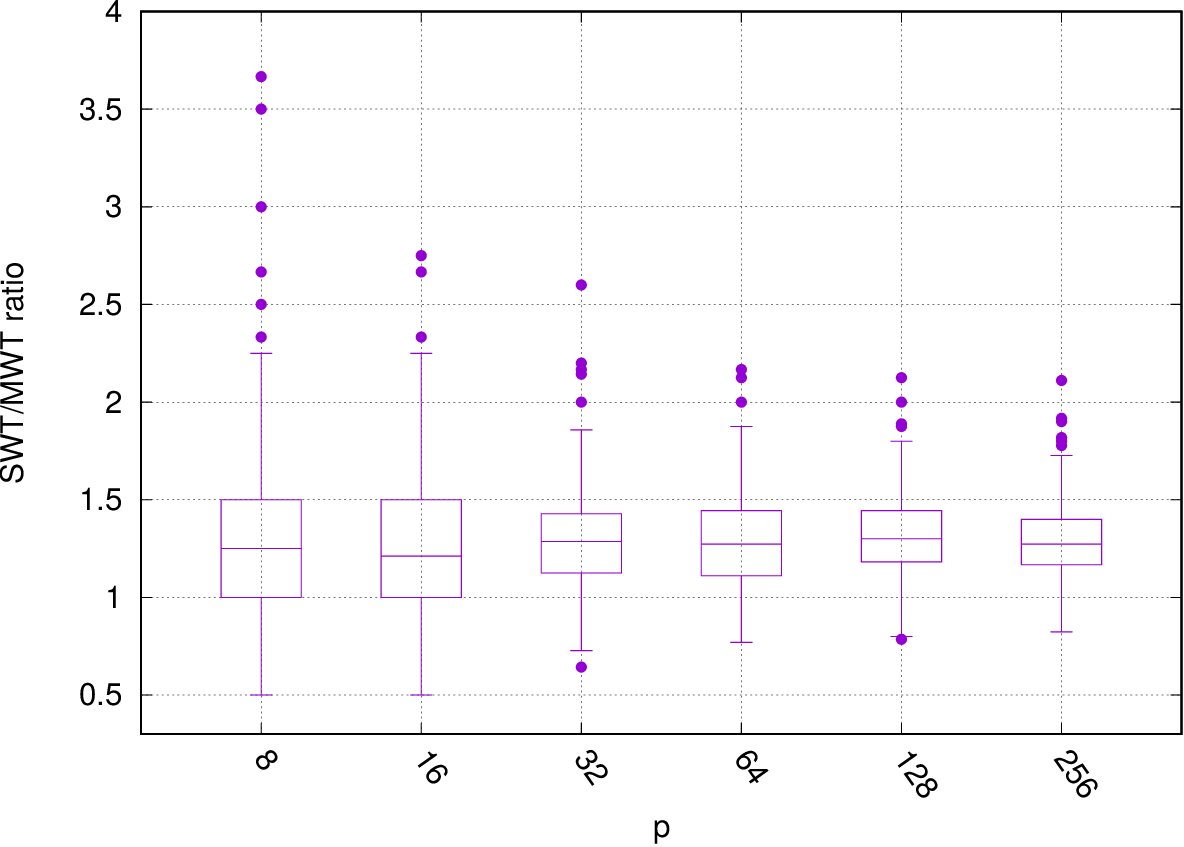}
	\caption{The ratio between the duration of the startup phase of the execution using \emph{MWT} and with \emph{SWT} according to the number of processors ($\lambda=262$ and $W=10^8$)} 
\label{fig:simultaneousefig}
\end{figure}
In practice, we observe that the \emph{MWT} mechanism only impacts significantly the startup phase.
Fig~\ref{fig:simulator:GantChartt} depicts an example of two scenarios which clarify the impact of \emph{MWT} and \emph{SWT} on the Gantt chart of the first phase.
As we see at time $t=0$, the processors $P_1$ and $P_2$ send to steal the processor $P_0$, and $P_3$ sends a steal request to $P_2$,
all steal requests arrive at the same time at $t=100$.
In the case of single work transfer \emph{SWT} in Fig~\ref{fig:simulator:GantChartt}-a, the processor $P_0$ answers with some of its work to $P_1$ and answers $P_2$ with failed responses.
At $t=200$, $P_1$ receives the stolen work and becomes active, and $P_2$ and $P_3$ try to steal again. Which is not the case in the case with multiple work transfer \emph{SWT} in Fig~\ref{fig:simulator:GantChartt}-b where the processor $P_0$ answer $P_1$ and $P_2$ at the same time. This act accelerates the increase in the number of active processors after each round trip (steal-answer), 
which is clear in the figure at $t=300$.

Fig~\ref{fig:simultaneousefig} presents in BoxPlot format the ratio between the duration of the startup phase using the \emph{MWT} mechanism and using the \emph{SWT} mechanism according to the number of processors. The x-axis is the number of processors and the y-axis is the ratio between the two durations of the startup phase using the \emph{SWT} and \emph{MWT} mechanisms for $\lambda=262$ and $W=10^8$.  
In this setting we see that \emph{MWT} is reducing the duration of the startup phase for 75\% of the runs with a gain larger than 200\% for a small number of processors.
The behavior of \emph{MWT} is positive on the startup phase but the overall performance gains are small because the duration of the startup phase is small compared to the total execution time. 

\section{Conclusion}
\label{sec:system:conclusion}

We present in this paper our lightweight PYTHON simulator for experimentally
analyze different model of \textit{Work Stealing} algorithms. Our simulator
is developed to be flexible enough to simulate different topologies and 
applications with different variants of \textit{Work Stealing} algorithms. 
Using this simulator, we provided and discussed the theoretical bound on 
the Makespan execution of \textit{Work Stealing} on one cluster topology 
founded in~\cite{Gast2018ANA}.
We also experimentally study the impact of simultaneous responses on one cluster.